\documentclass[preprint]{aastex631}

\graphicspath{{./}{figures/}}

\shorttitle{Comparison of magnetograms obtained with HMI/SDO and SP/Hinode}
\shortauthors{Zhang, Zhang \& Jiang}

\begin{document}

   \title{A comparison of co-temporal vector magnetograms obtained with HMI/SDO and SP/Hinode}

\author[0000-0002-3141-747X]{Mei Zhang}
\affiliation{National Astronomical Observatories, Chinese Academy of Sciences, Beijing 100101, China; zhangmei@nao.cas.cn}
\affiliation{School of Astronomy and Space Sciences, University of Chinese Academy of Sciences, Beijing 100049, China}
\affiliation{High Altitude Observatory, National Center for Atmospheric Research, 3080 Center Green Drive, Boulder, CO 80301, USA}

\author{Haocheng Zhang}
\affiliation{Jericho High School, 99 Cedar Swamp Road, Jericho, NY 11753, USA}

\author{Chengqing Jiang}
\affiliation{School of Space and Environment, Beihang University, Beijing 100091, China }

\begin{abstract}
An accurate measurement of magnetic field is very important for understanding the formation and evolution of solar magnetic fields. Currently there are two types of solar magnetic field measurement instruments: the filter-based magnetographs and the Stokes polarimeters. The former gives high temporal resolution magnetograms and the latter provides more accurate measurements of magnetic fields. Calibrating the magnetograms obtained by filter-based magnetographs with those obtained by Stokes polarimeters is a good way to combine the advantages of the two types. Our previous studies have shown that, compared to the magnetograms obtained by the Spectro-Polarimeter (SP) on board Hinode, those magnetograms obtained by both the filter-based Solar Magnetic Field Telescope (SMFT) of the Huairou Solar Observing Station (HSOS) and by the filter-based Michelson Doppler Imager (MDI) aboard SOHO have underestimated the flux densities in their magnetograms and systematic center-to-limb variations present in the magnetograms of both instruments. Here, using a sample of 75 vector magnetograms of stable alpha sunspots, we compare the vector magnetograms obtained by the Helioseismic and Magnetic Imager (HMI) aboard SDO with co-temporal vector magnetograms obtained by SP/Hinode. Our analysis shows that both the longitudinal and transverse flux densities in the HMI/SDO magnetograms are very close to those in the SP/Hinode magnetograms and the systematic center-to-limb variations in the HMI/SDO magnetograms are very minor. Our study suggests that using the filter-based magnetograph to construct a low spectral resolution Stokes profile, as done by HMI/SDO, can largely remove the disadvantages of the filter-type measurements and yet still possess the advantage of high temporal resolution.
\end{abstract}


\section{Introduction}        

It is well known that the Sun’s magnetic field plays an important role in controlling the solar activities such as coronal mass ejections \citep{Zhang2005}. Understanding how the solar magnetic fields are produced \citep{Charbonneau2014} and evolved is thus very crucial. For this purpose, an accurate measurement of the magnetic field \citep{Stenflo1994} with good spatial and temporal resolutions becomes vital.

The measurements of photospheric magnetic fields are mainly based on two types of solar magnetic field telescopes: the filter-based magnetographs and the Stokes polarimeters. They all use the Zeeman effect to measure the magnetic fields, but they take different approaches to achieve the measurements, making them possess their respective advantages and disadvantages.

A Stokes polarimeter measures the full spectra of Stokes I, Q, U, and V of a spectral line. An inversion code is applied to derive the vector magnetic field, together with other thermal parameters. Since the full spectra (with a high spectral resolution in most cases) is used in the inversion, the derived magnetic field is usually more accurate. Also, a parameter called the filling factor ($f$) can be obtained. This gives a more accurate measurement of the true field strength, particularly for the magnetic fields outside active regions where the filling factors are usually significantly less than 1. However, because the polarimeter needs to scan the investigated field of view step by step to get a magnetogram, the temporal resolution of the observation is usually low. For example, it could take the Spectro-Polarimeter (SP) on board Hinode 40 - 60 minutes to scan an area covering a typical active region.

A magnetograph measures the Stokes I, Q, U, and V maps at only one, or at most several, fixed wavelengths. Pre-calculated calibration coefficients or calibration maps are usually used to obtain the vector magnetograms. The advantage of the magnetograph measurements is their high temporal resolutions.  While it could take tens of minutes or even hours for a Stokes polarimeter to generate a piece of vector magnetogram of the solar active region, the typical value for a filter-based magnetograph to obtain a full-disk vector magnetogram is only a few minutes. However, an accurate calibration is not an easy undertaking \citep{SuJT2004} and many other parameters may come into play to influence the calibration. 

The difficulty on obtaining an accurate calibration for the filter-based magnetograph can be seen from the fact that the Michelson Doppler Imager (MDI on board SOHO) data has been recalibrated a few times. The original calibration \citep{Scherrer1995} used the standard center-of-gravity method. Later on, \citet{Berger2003} compared the MDI magnetograms with the co-temporal magnetograms obtained by the Advanced Stokes Polarimeter (ASP) and found that the calibration of the MDI magnetograms has underestimated the flux density by a factor about 1.6. Then based on a detailed cross-correlation between sets of magnetograms simultaneously obtained by the Mount Wilson Observatory (MWO) and by MDI/SOHO \citep{Tran2005}, the MDI team recalibrated all of the full-disk MDI magnetograms in October 2007 and referred these data as “version 2007 MDI level-1.8 data”. However, later on \citet{Ulrich2009} recommended a new calibration, which multiplies the previous calibration map \citep{Tran2005} by a factor that depends on the distance from the disk center.  This correction was applied to the MDI data in December 2008, which results in the “version 2008 MDI level-1.8 data.”
 
Since  \citet{Berger2003} first compared the MDI magnetograms with the co-temporal magnetograms obtained by the ASP to calibrate the MDI magnetograms, comparing the magnetograms obtained by filter-based magnetographs with those obtained by Stokes polarimeters to calibrate the filter-based magnetographs has become a popular approach.  On 2006 September 22, the Hinode satellite \citep{Kosugi2007} was launched. The Stokes polarimeter SP/Hinode began to provide possibly so-far the most accurate vector magnetograms. It is then wise to use SP/Hinode magnetograms to calibrate various filter-based magnetograph data. 

 \citet{WangD2009a} compared a set of co-temporal magnetograms obtained by the SP/Hinode with those obtained by the Solar Magnetic Field Telescope (SMFT) of the Huairou Solar Observing Station (HSOS) to check the linear calibrations of the SMFT vector magnetograms. They found that the used calibration coefficients of the SMFT \citep{SuJT2004} have under-estimated the flux density and meanwhile a strong center-to-limb variation of the calibration coefficients was not taken into account.  
 
Using the same approach, \citet{WangD2009b} compared a set of co-temporal magnetograms of active regions obtained by the MDI/SOHO and by the SP/Hinode. They found that, although the most recent calibration of the ``version 2008 MDI level-1.8 data'' has largely removed the center-to-limb variation that is severe in the ``version 2007 MDI level-1.8 data'', the magnetic flux density in the ``version 2008 MDI level-1.8 data'' is still lower than that in the SP/Hinode magnetograms. The average ratio between the ``version 2008 MDI level-1.8 data'' and the SP/Hinode magnetograms is 0.71, and is 0.82 for the ``version 2007 MDI level-1.8 data''. 

In early 2010, the Solar Dynamics Observatory (SDO) was launched. The Helioseismic and Magnetic Imager (HMI) on board SDO began to provide continuous observations of full-disk vector magnetograms from the space. In this paper, we carry out a cross calibration between the HMI/SDO and SP/Hinode vector magnetograms. The data and the sample will be described in Section 2. The analysis and results will be presented in Section 3. A brief conclusion and discussion will be given in Section 4.

\section{The data and the samples}

HMI \citep{Scherrer2012} aboard SDO is designed to study the magnetic fields of the photosphere and the oscillations of the Sun. Two 4096 $\times$ 4096 pixels CCD cameras on HMI provide us with full-polarimetric filtergrams at six carefully selected spectral points of the FeI 6173{\AA} line. With a spatial resolution of about 0.5$''$ $\times$ 0.5$''$ per pixel, observations at the six spectral points form a low spectral-resolution spectra at each pixel point. Unlike other filter-based magnetographs where the magnetograms are obtained by using pre-calculated calibration coefficients or calibration maps, the HMI vector magnetograms are obtained by using a Milne–Eddington based inversion code \citep{Borrero2011}, where the filling factor has been taken as 1. As we will see with the development of this paper, this inversion approach has successfully removed most disadvantages of the filter-type instruments caused by rough pre-calibrations.

For scientific investigations, HMI team provides a multitude of data products. In this paper, we use the $hmi.B\_720s$ data series. The ``$720s$'' means that a tapered temporal average is performed every 720 seconds using 360 filtergrams collected over a 1350-second interval. The vector magnetograms in this series are in the native coordinate, i.e. a 2D array as measured at each CCD pixel.  Since our study focuses on the field strengths of the longitudinal and transverse fields, the $180^{\circ}$ disambiguation solution becomes irrelevant, even though three different solutions have been provided by the HMI team.

SP/Hinode obtains line profiles of two magnetically sensitive Fe lines at 630.15 and 630.25 nm and nearby continuum, using a $0.16''\times164''$ slit. The SP/Hinode data are calibrated \citep{Lites2013} and inverted at the Community Spectro-polarimetric Analysis Center (CSAC, http://www.csac.hao.ucar.edu/). The inversion is based on the assumption of the Milne-Eddington atmosphere model and a nonlinear least-square fitting technique, where the analytical Stokes profiles are fitted to the observed profiles. The inversion gives 36 parameters including the three components of magnetic field and the filling factor. The resolution of the magnetograms is of either 0.16$''$/pixel  for normal maps or 0.32$''$/pixel  for fast maps. The durations of these maps are usually tens of minutes. 
 
 Our sample is made of 75 pairs of vector magnetograms, each pair consisting of one HMI magnetogram and one co-temporal SP magnetogram. They are of active regions of four alpha sunspots, that is, NOAA 11084, NOAA 11092, NOAA 11216 and NOAA 11582. The alpha sunspots are chosen because they are very stable, presenting very little evolution during their passages over the solar disk. Since the HMI magnetograms we use are taken at ``one-time'' (although integration time is 720 seconds) and yet it usually takes tens of minutes to scan the co-temporal SP magnetograms, using magnetograms of stable sunspots will obviously reduce the errors induced by the evolution of the sunspots. 
 
 \begin{figure}
\centering
\includegraphics[width=0.9\textwidth]{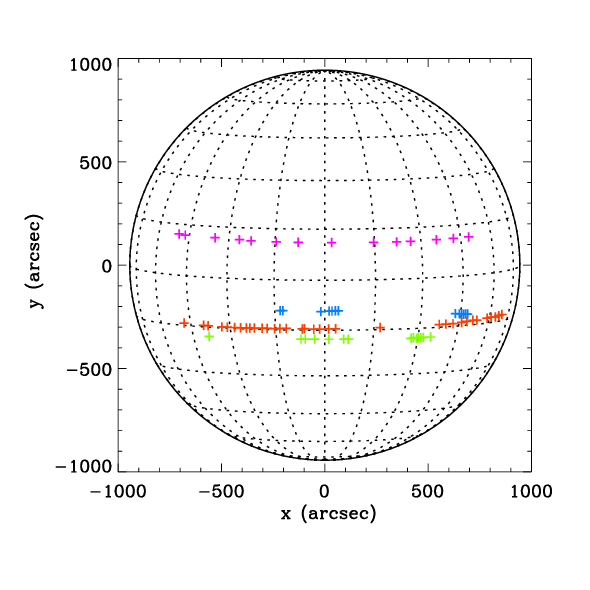}
\caption{Positions of the 75 magnetograms in the sample, with NOAA 11084, 11092, 11216 and 11582 plotting in green, purple, blue and red respectively.}
\label{fig1}
\end{figure} 

We first download all the SP magnetograms of these four sunspots. This gives 75 SP magnetograms. The on-disk positions of these 75 magnetograms are plotted in Figure \ref{fig1}, where different active regions have been presented with different colors. We can see that these magnetograms cover a wide range of longitudes on the solar disk, from near the solar limb to near the disk center. This is the reason why we choose these four active regions, for they have multiple observations from the center to the limb. 
 
After downloading these SP magnetograms, we read out the fits headers and calculate the middle times of each SP observations. We then go to the HMI webpage (http://hmi.stanford.edu/magnetic/) and download the full-disk vector magnetograms whose observation times are closest to the middle times of the SP magnetograms.  Information on the SP observation time periods, middle times of the SP observations and the HMI observation times of these 75-pair magnetograms are listed in Tables \ref{Tab1} and \ref{Tab2}. Also listed in Tables \ref{Tab1} and \ref{Tab2} are the latitudes ($\theta$), longitudes ($\phi$) and the heliocentric angles ($\rho$, $cos\rho=cos\theta \cdot cos\phi$) of the sunspots in these 75-pair magnetograms.

\section{Analysis and results}

To compare the 75-pair vector magnetograms in our sample, we first need to do an alignment and image scaling. This is to make each pair of the magnetograms have the same field of view and the same pixel size so that they can be compared pixel by pixel. 

\begin{figure}
\centering
\includegraphics[scale=1.0]{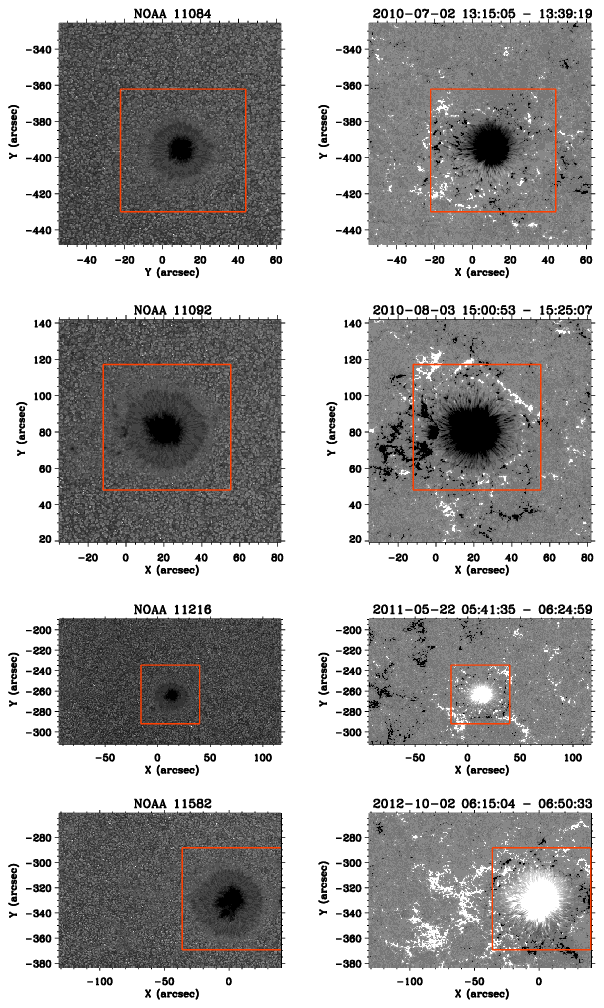}
\caption{Examples of the SP maps of the four active regions in the sample. Left panels are of the continuum intensity maps ($I_c$) and the right panels are of the longitudinal magnetograms ($B_L$). The red square in each panel outlines the region that we cut for the study in this paper. Presented on top of each left panel is the NOAA number and on top of each right panel is the date and the time of the SP observation.} 
\label{fig2}
\end{figure}

To do this, we first cut the field of view of each downloaded SP magnetograms. Examples are given in Figure \ref{fig2}. Presented from top to bottom in Figure \ref{fig2} are the continuum intensity maps ($I_c$, left panels) and the longitudinal magnetograms ($B_L$, right panels) of the four active regions, respectively. For each active region, we present the one that was taken at the location closest to the central meridian. Here for the SP data, we have  $B_{L} = f \cdot B \cos\phi$ and $B_{T} = \sqrt{f} \cdot B \sin\phi$, where $B$ is the inversion-derived field strength, $\phi$ is the field inclination and $f$ is the filling factor. Whereas each map in Figure \ref{fig2} shows the full field of view of the SP observation, the red square in each panel outlines the region that we cut for the study in this paper. We can see that the focus is on the sunspot regions and the networks have been excluded although not fully. 

After we cut the 75 SP $I_c$, $B_L$ and $B_T$ maps one by one, we then use the CONGRID function in IDL to reform these maps to have the same pixel size as that of the HMI maps. In the next step of alignment, we first use the gravity center of the SP $I_c$ map to get a rough position of the sunspot in the HMI full-disk $I_c$ map. Then we use a cross-correlation algorithm to overlay the reformed SP $B_L$ map on the HMI full-disk $B_L$ map and cut out the same field of view of the SP map to get the studied HMI $B_L$ map. The same alignment is then applied to the HMI full-disk $B_T$ map to cut out the studied HMI $B_T$ map. Note here for the HMI data,  $B_{L} = B \cos\phi$ and $B_{T} = B \sin\phi$, where $B$ is the inversion-derived field strength and $\phi$ is the field inclination. The filling factor $f$ is missing here because in the HMI inversion the filling factor $f$ has been set to 1. Also noteworthy is that the latitudes, longitudes  and the heliocentric angles of the sunspots that we present in Tables \ref{Tab1} and \ref{Tab2} are estimated by using the gravity center of the sunspot in the HMI full-disk $I_c$ map. We did not use the pointing information in the SP fits header because they can be incorrect by dozens of arcseconds, as pionted out in \citet{Fouhey2023}.

\begin{figure}
\centering
\includegraphics[scale=1.0]{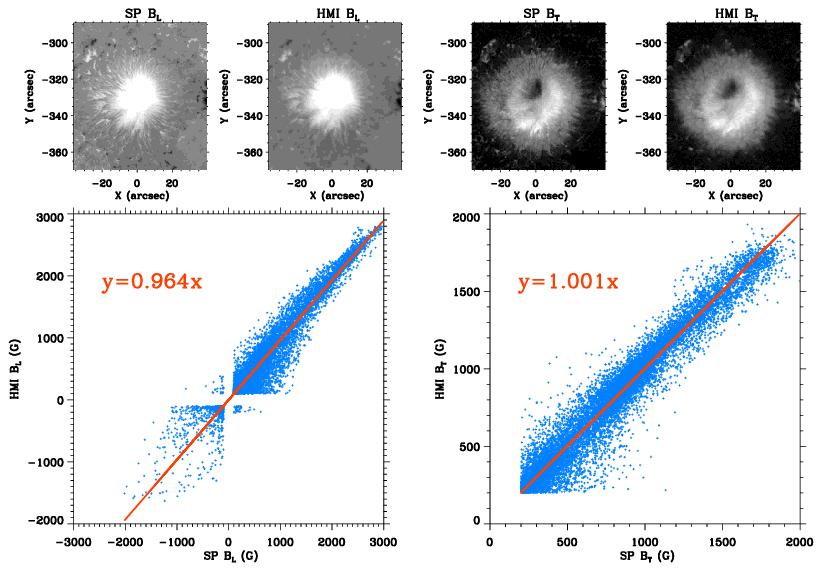}
\caption{Top panels: from left to right, SP $B_L$, HMI $B_L$, SP $B_T$ and HMI $B_T$ maps of NOAA 11582, when observed near the disk center (No. 61 pair in Table \ref{Tab2}). Bottom panels: The linear fitting between the SP $B_L$ and HMI $B_L$ maps (left), and the linear fitting between the SP $B_T$ and HMI $B_T$ maps (right). }
\label{fig3}
\end{figure}

\begin{figure}
\centering
\includegraphics[scale=1.0]{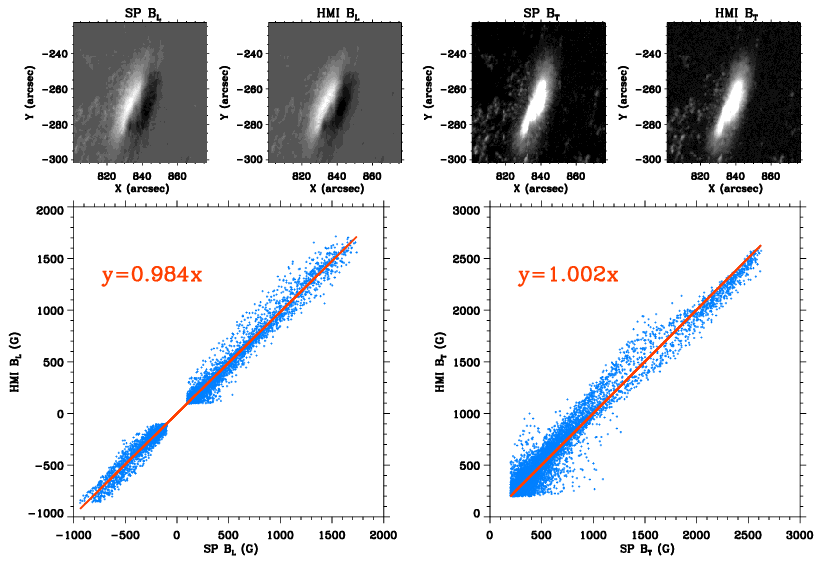}
\caption{Same as Figure \ref{fig3}, of the NOAA 11582, but when observed near the limb (No. 75 pair in Table \ref{Tab2}).}
\label{fig4}
\end{figure}

The results of the alignment and the image scaling can be seen from the examples in Figures \ref{fig3} and \ref{fig4}. On the top panels of Figure \ref{fig3} we show the SP $B_L$, HMI $B_L$, SP $B_T$ and HMI $B_T$ maps, from the left to right panel respectively, of the active region NOAA 11582, when it was observed near the disk center. This is the No. 61 pair in the Table \ref{Tab2}. We can see that the alignment works well. This good quality is also evident in the top panels of Figure \ref{fig4}, where the SP $B_L$, HMI $B_L$, SP $B_T$ and HMI $B_T$ maps, from the left to right panel respectively, are shown, again of the active region NOAA 11582, but when it was observed near the solar limb. This pair, presented in Figure \ref{fig4}, is the No.75 pair in the Table \ref{Tab2}.

After the successful alignment and image scaling, we then do a linear fitting between the SP $B_L$ data points and the HMI $B_L$ data points. An example is given in the left bottom panel of Figure \ref{fig3}, for the No. 61 pair. The blue plus symbols here are of the $B_L$ values, x-axis of the SP values and y-axis of the HMI values. The red thick line is the result of a linear fitting, $y=R_L\cdot x$, where data points with $|B_L|<100 G$ have been excluded. The fitting gives $y=0.964 ~x$, which means that the flux density in the HMI $B_L$ map is about 96.4\% of that of the SP $B_L$ map. This is already very close to 1. Note that on average the flux density in the version 2008 MDI data is only 71\% of that of the SP \citep{WangD2009b}. 

In a similar way, we carry out a linear fitting between the SP $B_T$ data points and the HMI $B_T$ data points. The right bottom panel of Figure \ref{fig3} shows the result of the linear fitting between the SP $B_T$ map and the HMI $B_T$ map. The blue plus symbols here are of the $B_T$ values, x-axis indicates the SP values and y-axis the HMI values. The red thick line is the result of a linear fitting, $y=R_T\cdot x$, where data points with $|B_T|<200 G$ (about $2 \sigma$ noise level for transverse fields) have been excluded. We see here that the fitting gives $y=1.001 ~x$, which means that the flux density in the HMI $B_T$ map is very close to that in the SP $B_T$ map, closer than the HMI $B_L$ map with respective to the SP $B_L$ map. 

In a same way, the bottom panels in Figure \ref{fig4} show the fitting results for the No.75 pair. For this pair,  $R_L=0.984$ and $R_T=1.002$. We see the same trend that both the flux densities in the HMI $B_L$ and $B_T$ maps are very close to those in the SP maps.  This means that the problem of the under-estimation of the flux density in previous SMFT and MDI data has been largely overcome.

\begin{figure}
\centering
\includegraphics[scale=1.0]{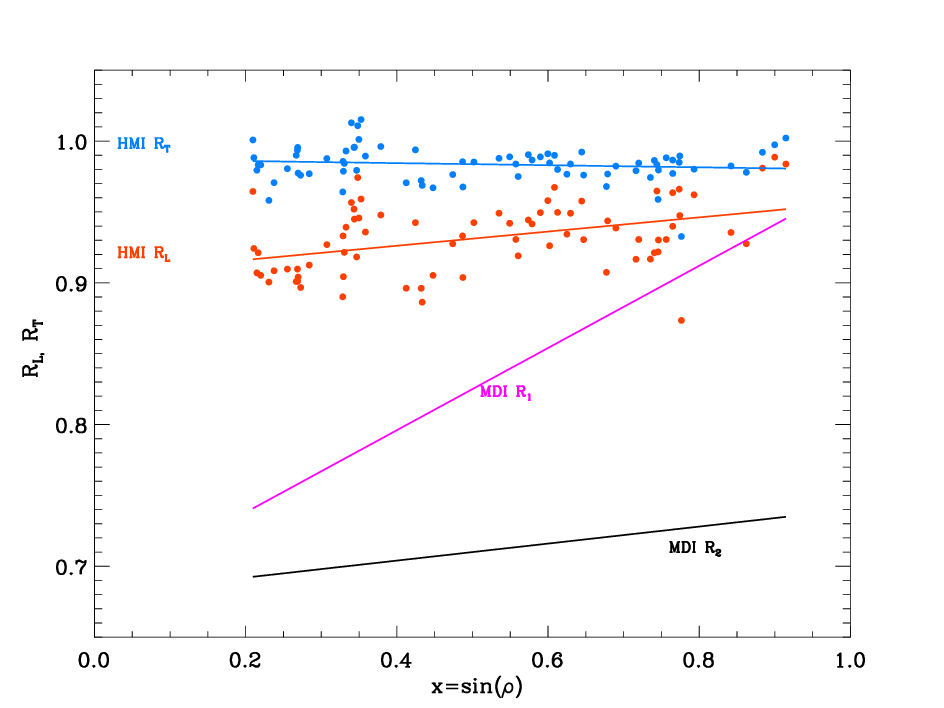}
\caption{Variations of HMI $R_L$ (red filled-circles) and HMI $R_T$ (blue filled-circles) with the heliocentric angle ($\rho$). The solid lines show the results of a linear fitting. They are $y = 0.91 + 0.05 x$ (red line for  $R_L$) and $y = 0.99 - 0.007 x$ (blue line for $R_T$), where $x=sin (\rho)$. The purple and black lines show the fitting results of the MDI version 2007 and version 2008 respectively. They are $y = 0.68 + 0.29 x$ (purple line) and $y = 0.68 + 0.06 x$ (black line), taken from \citet{WangD2009b}.} 
\label{fig5}
\end{figure}

The fitting results of the 75 $R_L$ values and 75 $R_T$ values are listed in Tables \ref{Tab1} and \ref{Tab2}, as well as plotted out in Figure \ref{fig5}. Red filled-circles in Figure \ref{fig5} show the 75 $R_L$ values and blue filled-circles show the 75 $R_T$ values. The heliocentric angle ($\rho$) is indicated along x-axis. The solid red line shows the result of a linear fitting of the 75 $R_L$ values with the x-axis, where $x = sin (\rho)$. The result gives $R_L = 0.91 + 0.05 x$. The solid blue line shows the result of a linear fitting of the 75 $R_T$ values with $x = sin (\rho)$. The result gives $R_T = 0.99 - 0.007 x$.  We see here that all the $R_L$ and $R_T$ values are close to 1, and the center-to-limb variation (that is, the dependence on x) is not large. The mean value of  $R_L$ is 0.932 and the mean value of $R_T$ is 0.984. 

As a comparison, the average ratio between the ``version 2008 MDI level-1.8 data'' and the SP/Hinode magnetograms is 0.71, and 0.82 for the ``version 2007 MDI level-1.8 data''. A similar linear fitting \citep{WangD2009b} gives the results as $y = 0.68 + 0.06 x$ for the ``version 2008 MDI level-1.8 data'' and $y = 0.68 + 0.29 x$ for the ``version 2007 MDI level-1.8 data''. These two fitting results of the MDI data are also plotted out in Figure \ref{fig5}, as purple (2007 version) and black (2008 version) lines. By comparing the four solid lines, we can see that indeed the HMI data have largely overcome the problems of both the under-estimation of the flux density and the existence of a center-to-limb variation that present in previous MDI magnetograms.

\section{Conclusion and discussion}

In this paper we compared a set of co-temporal alpha sunspots magnetograms obtained by HMI/SDO and by SP/Hinode. A pixel-by-pixel comparison of the magnetograms shows that the flux density in the HMI/SDO longitudinal magnetograms is about 0.93 of that of the SP/Hinode, and the flux density in the HMI/SDO transverse magnetograms is about 0.98 of that of the SP/Hinode. Moreover, the center-to-limb variation, which was severe in previous filter-type magnetograph data, is very minor now. We conclude that the HMI/SDO data have largely overcome the problems found in previous filter-type data, namely, an under-estimation of the flux density and a severe center-to-limb variation. Our investigation indicates that, using the filter-based magnetograph to scan at a few spectral points to form a low spectral resolution Stokes profile, as that in the HMI/SDO, can largely remove the disadvantages of the filter-type magnetograph measurement, and yet still possess high temporal resolution as the advantage.

With this conclusion, a few clarifications are in order. First, there is a limitation that low spectral resolution measurements cannot deal with. That is, in the complicated configurations of the magnetic fields, the complicated circular polarization profiles with central reversal cannot be found by the low spectral resolution observations like HMI/SDO. Second, with large-aperture telescopes such as DKIST and EST coming, the temporal resolution of magnetograms obtained by Stokes Polarimeters are getting better and better. Finally, it needs to be pointed out that these two kinds of instruments share the shortage that the 3D (two dimensions of space plus one dimension of dispersion) polarimetric data  cannot be obtained simultaneously. The real time 3D polarimetric data will be obtained by the future instruments based on integral field units (IFUs) like those to be mounted in FASOT \citep{Qu2011, Qu2017, Qu2022}.

\begin{acknowledgements}
We thank the referee for helpful comments and suggestions. The HMI data used in this paper were provided by courtesy of NASA/SDO and the HMI science team. Hinode is a Japanese mission developed and launched by ISAS/JAXA, collaborating with NAOJ as a domestic partner, NASA and STFC (UK) as international partners. The Hinode SOT/SP data used in this paper were distributed by the Community Spectropolarimetric Analysis Center of HAO/NCAR.  This work is supported by the National Natural Science Foundation of China (grant No. 11973056) and the National Key R\&D Program of China (grant No. 2021YFA1600500).
\end{acknowledgements}

\begin{table}
\scriptsize
\caption{Sample information and results}
\label{Tab1}
\begin{tabular}{llllllllllll}
\hline
No. & NOAA  & Date       & SP Time             & Middle time & HMI Time & Latitude   & Longitude   & $\rho $     & $R_L$   & $R_T$   \\
    &       &            &        (UT)              &    (UT)         &  (UT)        & (degrees)   & (degrees)   & (degrees)      &    &    \\\hline
1   & 11084 & 2010/06/29 & 11:20:00 - 11:44:36  & 11:32:29 & 11:34:26 & -19.3 & -38.8 & 42.6  & 0.907 & 0.968 \\
2   & 11084 & 2010/07/01 & 21:15:40 - 21:39:54  & 21:27:47 & 21:34:26 & -19.3 & -7.4  & 20.6  & 0.959 & 1.015 \\
3   & 11084 & 2010/07/01 & 23:55:50 - 00:20:04 & 00:07:57 & 23:58:26 & -19.2 & -6.1  & 20.1  & 0.945 & 0.996 \\
4   & 11084 & 2010/07/02 & 05:15:05 - 05:39:19 & 05:27:12 & 05:34:26 & -19.2 & -3.1  & 19.4    & 0.939 & 0.993 \\
5   & 11084 & 2010/07/02 & 13:15:05 - 13:39:19 & 13:27:12 & 13:34:26 & -19.3 & 1.3  & 19.3  & 0.921 & 0.984 \\
6   & 11084 & 2010/07/02 & 21:50:05 - 22:14:19 & 22:02:12 & 21:58:26 & -19.2 & 5.8   & 20.1   & 0.952 & 0.996 \\
7   & 11084 & 2010/07/03 & 00:32:50 - 00:57:04 & 00:44:57 & 00:46:26 & -19.2 & 7.3   & 20.5   & 0.946 & 1.001 \\
8   & 11084 & 2010/07/04 & 14:50:53 - 15:15:08 & 15:03:01 & 14:58:26 & -19.1 & 27.8  & 33.3  & 0.942 & 0.989 \\
9   & 11084 & 2010/07/04 & 16:20:05 - 16:44:19 & 16:32:12 & 16:34:26 & -19.0 & 28.6  & 33.9    & 0.931 & 0.984 \\
10  & 11084 & 2010/07/04 & 18:14:35 - 18:38:27 & 18:26:01 & 18:34:26 & -19.2 & 29.8  & 35.0    & 0.944 & 0.990 \\
11  & 11084 & 2010/07/04 & 19:51:36 - 20:14:48 & 20:02:42 & 19:58:26 & -19.0 & 30.4  & 35.4   & 0.942 & 0.987 \\
12  & 11084 & 2010/07/04 & 21:26:32 - 21:50:46 & 21:38:39 & 21:34:26 & -19.0 & 31.3  & 36.1    & 0.949 & 0.989 \\
13  & 11084 & 2010/07/04 & 22:55:05 - 23:19:19 & 23:07:12 & 23:10:26 & -19.0 & 32.2  & 36.8    & 0.958 & 0.991 \\
14  & 11084 & 2010/07/05 & 03:50:05 - 04:14:20 & 04:02:13 & 03:58:26 & -18.9 & 34.8  & 39.0    & 0.949 & 0.984 \\
15  & 11092 & 2010/07/30 & 17:26:49 - 17:56:56 & 17:41:53 & 17:34:26 & 12.8  & -49.7 & 50.9  & 0.873 & 0.933 \\
16  & 11092 & 2010/07/30 & 22:22:40 - 23:21:56 & 22:51:48 & 22:46:26 & 12.7  & -46.9 & 48.2  & 0.921 & 0.959 \\
17  & 11092 & 2010/07/31 & 20:01:40 - 20:25:53 & 20:13:47 & 20:10:25 & 12.7  & -35.0 & 37.0   & 0.926 & 0.985 \\
18  & 11092 & 2010/08/01 & 10:23:30 - 10:40:35 & 10:31:33 & 10:34:26 & 12.7  & -26.5 & 29.2    & 0.904 & 0.968 \\
19  & 11092 & 2010/08/01 & 20:30:51 - 20:55:05 & 20:42:58 & 18:46:25 & 12.5  & -22.6 & 25.7    & 0.886 & 0.969 \\
20  & 11092 & 2010/08/02 & 09:14:04 - 09:38:18 & 09:26:11 & 09:22:26 & 12.5  & -14.7 & 19.2   & 0.890 & 0.964 \\
21  & 11092 & 2010/08/02 & 21:00:49 - 21:25:02 & 21:12:56 & 21:10:25 & 12.5  & -8.0  & 14.8    & 0.910 & 0.981 \\
22  & 11092 & 2010/08/03 & 15:00:53 - 15:25:07 & 15:13:00 & 15:10:25 & 12.5 & 2.0  & 12.7 & 0.905 & 0.983 \\
23  & 11092 & 2010/08/04 & 14:00:50 - 14:25:05 & 14:12:58 & 14:10:25 & 12.5 & 14.7  & 19.2  & 0.904 & 0.979 \\
24  & 11092 & 2010/08/05 & 03:00:05 - 03:24:19 & 03:12:12 & 03:10:25 & 12.4  & 22.0  & 25.1  & 0.942 & 0.994 \\
25  & 11092 & 2010/08/05 & 11:09:06 - 11:33:21 & 11:21:14 & 11:22:25 & 12.4 & 26.6  & 29.1  & 0.933 & 0.985 \\
26  & 11092 & 2010/08/06 & 03:28:24 - 03:52:38 & 03:40:31 & 03:34:25 & 12.5  & 35.6  & 37.5  & 0.967 & 0.990 \\
27  & 11092 & 2010/08/06 & 15:15:51 - 15:40:05 & 15:27:58 & 15:22:25 & 12.4  & 42.1  & 43.6  & 0.939 & 0.982 \\
28  & 11092 & 2010/08/07 & 03:01:05 - 03:25:19 & 03:13:12 & 03:10:25 & 12.4  & 48.7  & 49.9  & 0.964 & 0.987 \\
29  & 11216 & 2011/05/21 & 01:57:40 - 02:21:54 & 02:09:47 & 02:10:24 & -15.3 & -13.6 & 20.4    & 0.974 & 1.011 \\
30  & 11216 & 2011/05/21 & 03:35:05 - 03:59:19 & 03:47:12 & 03:46:24 & -15.3 & -12.8 & 19.9    & 0.957 & 1.013 \\
31  & 11216 & 2011/05/22 & 00:46:05 - 01:29:29 & 01:07:17 & 01:10:24 & -15.6 & -1.2  & 15.6   & 0.901 & 0.996 \\
32  & 11216 & 2011/05/22 & 05:41:35 - 06:24:59 & 06:02:47 & 05:58:24 & -15.4 & 1.3   & 15.5   & 0.901 & 0.990 \\
33  & 11216 & 2011/05/22 & 07:20:05 - 08:03:28 & 07:41:17 & 07:34:24 & -15.4 & 2.2   & 15.6   & 0.910 & 0.994 \\
34  & 11216 & 2011/05/22 & 08:58:35 - 09:41:59 & 09:19:47 & 09:22:24 & -15.3 & 3.2   & 15.6   & 0.904 & 0.977 \\
35  & 11216 & 2011/05/22 & 10:37:05 - 11:20:28 & 10:58:17 & 10:58:24 & -15.3 & 4.1   & 15.8  & 0.900 & 0.976 \\
36  & 11216 & 2011/05/25 & 11:00:06 - 11:43:30 & 11:21:18 & 11:22:24 & -15.4 & 43.6  & 45.8   & 0.917 & 0.979 \\
37  & 11216 & 2011/05/25 & 14:07:35 - 14:50:59 & 14:28:47 & 14:34:24 & -15.5 & 45.3  & 47.3   & 0.917 & 0.974 \\
38  & 11216 & 2011/05/25 & 16:00:05 - 16:43:30 & 16:21:18 & 16:22:24 & -15.4 & 46.3  & 48.3   & 0.930 & 0.980 \\
39  & 11216 & 2011/05/25 & 17:40:05 - 18:23:28 & 18:01:17 & 17:58:24 & -15.5 & 47.2  & 49.1  & 0.931 & 0.988 \\
40  & 11216 & 2011/05/25 & 19:15:05 - 19:58:29 & 19:36:17 & 19:34:24 & -15.5 & 48.1  & 49.9   & 0.940 & 0.977 \\
41  & 11216 & 2011/05/25 & 20:50:05 - 21:33:29 & 21:11:17 & 21:10:24 & -15.5 & 49.0 & 50.7   & 0.947 & 0.989 \\
\hline
\end{tabular}
\end{table}

\begin{table}
\scriptsize
\caption{Table 1 continued}
\label{Tab2}
\begin{tabular}{llllllllllll}
\hline
No. & NOAA  & Date       & SP Time             & Middle time & HMI Time & Latitude   & Longitude   & $\rho $     & $R_L$   & $R_T$   \\
     &       &            &        (UT)              &    (UT)         &  (UT)        & (degrees)   & (degrees)   & (degrees)      &    &    \\\hline
42  & 11582 & 2012/09/28 & 16:00:48 - 16:33:08 & 16:16:58 & 16:10:18 & -12.3 & -46.6 & 47.8  & 0.921 & 0.986 \\
43  & 11582 & 2012/09/29 & 05:57:06 - 06:29:26 & 06:13:16 & 06:10:18 & -12.4 & -38.7 & 40.3  & 0.930 & 0.976 \\
44  & 11582 & 2012/09/29 & 09:05:04 - 09:37:24 & 09:21:14 & 09:22:18 & -12.4 & -36.9 & 38.7   & 0.934 & 0.977 \\
45  & 11582 & 2012/09/29 & 18:08:57 - 18:44:26 & 18:26:12 & 18:22:18 & -12.4 & -32.0 & 34.1  & 0.919 & 0.975 \\
46  & 11582 & 2012/09/29 & 21:30:05 - 22:05:35 & 21:47:20 & 21:46:18 & -12.3 & -30.1 & 32.4    & 0.949 & 0.988 \\
47  & 11582 & 2012/09/30 & 02:00:05 - 02:35:35 & 02:17:20 & 02:10:18 & -12.3 & -27.7 & 30.1    & 0.942 & 0.985 \\
48  & 11582 & 2012/09/30 & 05:30:05 - 06:05:34 & 05:47:20 & 05:46:18 & -12.3 & -25.6 & 28.3    & 0.928 & 0.976 \\
49  & 11582 & 2012/09/30 & 08:50:05 - 09:25:34 & 09:07:20 & 09:10:18 & -12.2 & -23.8 & 26.6   & 0.905 & 0.967 \\
50  & 11582 & 2012/09/30 & 10:59:05 - 11:34:35 & 11:16:20 & 11:10:18 & -12.2 & -22.7 & 25.6  & 0.896 & 0.972 \\
51  & 11582 & 2012/09/30 & 13:30:32 - 14:05:34 & 13:47:33 & 13:46:18 & -12.2 & -21.3 & 24.4    & 0.896 & 0.971 \\
52  & 11582 & 2012/09/30 & 17:56:56 - 18:32:25 & 18:14:11 & 18:10:18 & -12.1 & -18.8 & 22.3  & 0.948 & 0.996 \\
53  & 11582 & 2012/09/30 & 20:40:05 - 21:15:35 & 20:57:20 & 20:58:18 & -12.1 & -17.2 & 21.0   & 0.936 & 0.989 \\
54  & 11582 & 2012/10/01 & 01:00:05 - 01:35:34 & 01:17:20 & 01:10:18 & -12.1 & -15.0 & 19.2   & 0.933 & 0.986 \\
55  & 11582 & 2012/10/01 & 04:00:04 - 04:35:34 & 04:17:19 & 04:10:18 & -12.1 & -13.3 & 17.9    & 0.927 & 0.988 \\
56  & 11582 & 2012/10/01 & 07:20:04 - 07:55:34 & 07:37:19 & 07:34:18 & -12.0 & -11.4 & 16.5    & 0.912 & 0.977 \\
57  & 11582 & 2012/10/01 & 16:10:03 - 16:45:28 & 16:27:16 & 16:22:17 & -12.1 & -6.7  & 13.7  & 0.908 & 0.971 \\
58  & 11582 & 2012/10/01 & 19:20:04 - 19:55:34 & 19:37:19 & 17:34:17 & -12.0 & -6.0  & 13.3   & 0.901 & 0.958 \\
59  & 11582 & 2012/10/01 & 22:30:04 - 23:05:33 & 22:47:19 & 22:46:17 & -12.1 & -3.3  & 12.5  & 0.921 & 0.983 \\
60  & 11582 & 2012/10/02 & 01:30:04 - 02:05:34 & 01:47:19 & 01:46:18 & -12.1 & -1.5  & 12.1  & 0.924 & 0.988 \\
61  & 11582 & 2012/10/02 & 06:15:04 - 06:50:33 & 06:32:19 & 06:34:18 & -12.0 & 1.1   & 12.1   & 0.964 & 1.001 \\
62  & 11582 & 2012/10/02 & 09:53:06 - 10:28:36 & 10:10:21 & 10:10:17 & -12.0 & 3.2   & 12.4   & 0.907 & 0.979 \\
63  & 11582 & 2012/10/03 & 10:15:00 - 10:50:28 & 10:32:14 & 10:34:17 & -11.9 & 16.5  & 20.3   & 0.918 & 0.979 \\
64  & 11582 & 2012/10/04 & 21:15:00 - 21:50:29 & 21:32:15 & 21:34:17 & -12.1 & 36.0  & 37.8   & 0.950 & 0.980 \\
65  & 11582 & 2012/10/05 & 02:00:05 - 02:35:34 & 02:17:20 & 02:10:17 & -12.2 & 38.5  & 40.1   & 0.958 & 0.992 \\
66  & 11582 & 2012/10/05 & 07:00:05 - 07:35:34 & 07:17:20 & 07:10:17 & -12.3 & 41.3  & 42.7  & 0.944 & 0.977 \\
67  & 11582 & 2012/10/05 & 13:15:11 - 13:50:41 & 13:32:26 & 13:34:17 & -12.2 & 44.8  & 46.1   & 0.931 & 0.985 \\
68  & 11582 & 2012/10/05 & 17:00:04 - 17:29:55 & 17:14:30 & 17:10:17 & -12.0 & 46.9  & 48.1    & 0.965 & 0.983 \\
69  & 11582 & 2012/10/05 & 21:34:00 - 22:09:29 & 21:51:15 & 21:58:17 & -12.0 & 49.6  & 50.7  & 0.966 & 0.985 \\
70  & 11582 & 2012/10/06 & 01:00:05 - 01:35:35 & 01:17:20 & 01:22:17 & -12.1 & 51.5 & 52.5   & 0.962 & 0.980 \\
71  & 11582 & 2012/10/06 & 10:34:50 - 10:59:04 & 10:46:57 & 10:46:17 & -12.0 & 56.5  & 57.4  & 0.935 & 0.982 \\
72  & 11582 & 2012/10/06 & 14:30:06 - 14:54:19 & 14:42:13 & 14:46:17 & -11.9 & 58.8  & 59.6  & 0.928 & 0.978 \\
73  & 11582 & 2012/10/06 & 19:00:05 - 19:24:19 & 19:12:12 & 19:10:17 & -12.1 & 61.4 & 62.1 & 0.981 & 0.992 \\
74  & 11582 & 2012/10/06 & 23:00:05 - 23:24:20 & 23:12:13 & 23:10:17 & -11.9 & 63.5  & 64.1  & 0.989 & 0.997 \\
75  & 11582 & 2012/10/07 & 03:00:05 - 03:24:19 & 03:12:12 & 03:10:17 & -11.8 & 65.6  & 66.2  & 0.984 & 1.002 \\
\hline
\end{tabular}
\end{table}

\bibliography{ms}{}
\bibliographystyle{aasjournal}

\end{document}